\documentclass[a4paper,man,natbib,floatsintext]{apa6}
\usepackage{apacite}
\usepackage{subfig}
\usepackage[english]{babel}
\usepackage[utf8x]{inputenc}
\usepackage{graphicx}
\usepackage[colorinlistoftodos]{todonotes}
\usepackage[left]{lineno}
\usepackage{diagbox}
\usepackage{indentfirst}
\usepackage{adjustbox}
\usepackage{color}
\usepackage{soul}
\usepackage{amsmath,amssymb,amsfonts}
\usepackage{algorithm}
\usepackage{algpseudocode}
\usepackage{textcomp}
\usepackage{xcolor}
\usepackage{subfig}
\usepackage{caption}
\DeclareCaptionLabelFormat{andtable}{#1~#2  \&  \tablename~\thetable}
\usepackage{hyperref}
\usepackage{wrapfig}

\begin{document}

\raggedbottom

 \captionsetup[table]{
  labelsep = newline,
  textfont = sc, 
  name = TABLE, 
  justification=raggedleft,
  singlelinecheck=off,
  labelsep=colon,
  skip = \medskipamount}

\begin{titlepage}

\begin{center}

\large Enhancing autonomy transparency: an option-centric rationale approach\\ 

\normalsize

\vspace{25pt}
Ruikun Luo\\
Robotics Institute, University of Michigan, Ann Arbor\\
\vspace{15pt}
Na Du \\
Industrial and Operations Engineering, University of Michigan, Ann Arbor\\
\vspace{15pt}
X. Jessie Yang \\
Industrial and Operations Engineering, University of Michigan, Ann Arbor\\
\vspace{15pt}

\end{center}
\begin{flushleft}
\vspace{30pt}
\textbf{Manuscript type:} \textit{Original Research}\\ 
\textbf{Running head:} Enhancing autonomy transparency\\
\textbf{Word count:} 4200\\

\textbf{Corresponding author:} X. Jessie Yang, 1205 Beal Avenue, Ann Arbor, MI48109, Email: xijyang@umich.edu\\

\textbf{Acknowledgement:} We would like to thank Kevin Y. Huang for his assistance in data collection.
\end{flushleft}

\end{titlepage}
\shorttitle{}

\section{Abstract}

While the advances in artificial intelligence and machine learning empower a new generation of autonomous systems for assisting human performance, one major concern arises from the human factors perspective:  Humans have difficulty deciphering autonomy-generated solutions and increasingly perceive autonomy as a mysterious black box. The lack of transparency contributes to the lack of trust in autonomy and sub-optimal team performance. To enhance autonomy transparency, this study proposed an \textit{option-centric rationale} display and evaluated its effectiveness. We developed a game \textit{Treasure Hunter} wherein a human uncovers a map for treasures with the help from an intelligent assistant, and conducted a human-in-the-loop experiment with 34 participants. Results indicated that by conveying the intelligent assistant's decision-making rationale via the \textit{option-centric rationale} display, participants had higher trust in the system and calibrated their trust faster. Additionally, higher trust led to higher acceptance of recommendations from the intelligent assistant, and in turn higher task performance.  

\textbf{Keyword:} Transparent autonomy, Transparent automation, Design rationale, Trust in automation, Trust calibration, Propositional logic. 










\newpage

\section{1. INTRODUCTION}

While the advances in artificial intelligence and machine learning empowers a new generation of autonomous systems for assisting human performance, one major concern arises from the human factors perspective:  Human agents have difficulty deciphering autonomy-generated solutions and increasingly perceived autonomy as a mysterious black box. The lack of transparency contributes to the lack of trust in autonomy and sub-optimal team performance \citep{Chen:2014cmba,Endsley:2017,Lyons:2014,deVisser:2018ju, Yang:2017:EEU:2909824.3020230, Lyons:2016co, Du2019TRC}.

There are multiple definitions of autonomy transparency, to name a few: 
``the [degree of] shared intent and shared awareness between a human and a machine \citep{Lyons:2014}'', ``the extent to which an autonomous agent can convey its intent, performance, future plans and reasoning process \citep{Chenetal:2014}'', ``a mechanism to expose the decision-making of a robot \citep{Theodoreu:2017}'', ``the understandability and predictability of their actions \citep{Endsley:2017}'', ``the ability for the automation to be inspectable or viewable in the sense that its mechanisms and rationale can be readily known \citep{Miller:2018}''. Despite the lack of a universal definition, a fairly consistent pattern can be observed: a transparent autonomy should communicate to the human agent the autonomy's ability and performance, its decision-making logic and rationale, and its intent and future plans. 

Although autonomy transparency was only recently defined, research has been conducted to convey certain aspects of autonomy-generated solutions. One body of human factors research has concentrated on conveying likelihood information in the form of automation reliability, (un)certainty, and confidence. Some studies revealed that likelihood information significantly helped human operators calibrate their trust and enhance human-automation team performance \citep{mcguirl2006supporting, walliser2016application, Wang2009Trust}. Other studies reported that human operators did not trust or depend on automated decision aids appropriately even when the likelihood information was disclosed \citep{Bagheri2004, fletcher2017visualizing}. Recently, \cite{Du2019} proposed a framework for reconcile the mixed results and showed that not all likelihood information is equal in aiding human-autonomy team performance. Presenting the predictive values and the overall success likelihood is more beneficial than presenting the hit and correct rejection rates. 

Another body of research has investigated the impact of providing hand-crafted explanations of autonomy’s behaviors. For example,  \cite{Dzindolet:2003kjba} showed that providing hand-drafted explanations of automation failures can lead to increased trust. The studies of \citet{Koo:2014eg} and \citet{Koo:2016hx} showed that by informing the drivers of the reasons for automated breaking (e.g. road hazard ahead) decreased drivers’ anxiety, increased their sense of control, preference and acceptance. Similarly, \cite{Du2019TRC} found that speech output explaining why and how the automated vehicle is going to take certain actions was rated higher on trust, preference, usability and acceptance.

More recently, research has gone beyond either conveying likelihood information or relying on hand-crafted explanation, and has formally defined and examined autonomy transparency. Notably, \cite{Mercado2016Intelligent} proposed the situation awareness (SA)-based agent transparency model to convey information supporting the human agent’s perception, comprehension, and projection of an intelligent assistant's recommendations.

In this study, we wished to propose the \textit{option-centric rationale} approach for enhancing autonomy transparency. Inspired by the research on design rationale, the option-centric rationale approach explicitly displays the option space (i.e. all the possible options/actions that an autonomy could take) and the rationale why a particular option is the most appropriate at a given context. 
Design rationale is an area of design science focusing on the ``representation for explicitly documenting the reasoning and argumentation that make sense of a specific artifact \citep{MacLean:1991:QOC:1456153.1456155}''. Its primary goal is to support designers and other stake holders by recording the argumentation and reasoning behind the design process. The theoretical underpinning for design rationale is that for designers what is important is not just the specific artifact itself but its other possibilities -- why an artifact is designed in a particular way compared to how it might otherwise be.

\begin{figure} [h]
    \centering
    \includegraphics[width=1\textwidth]{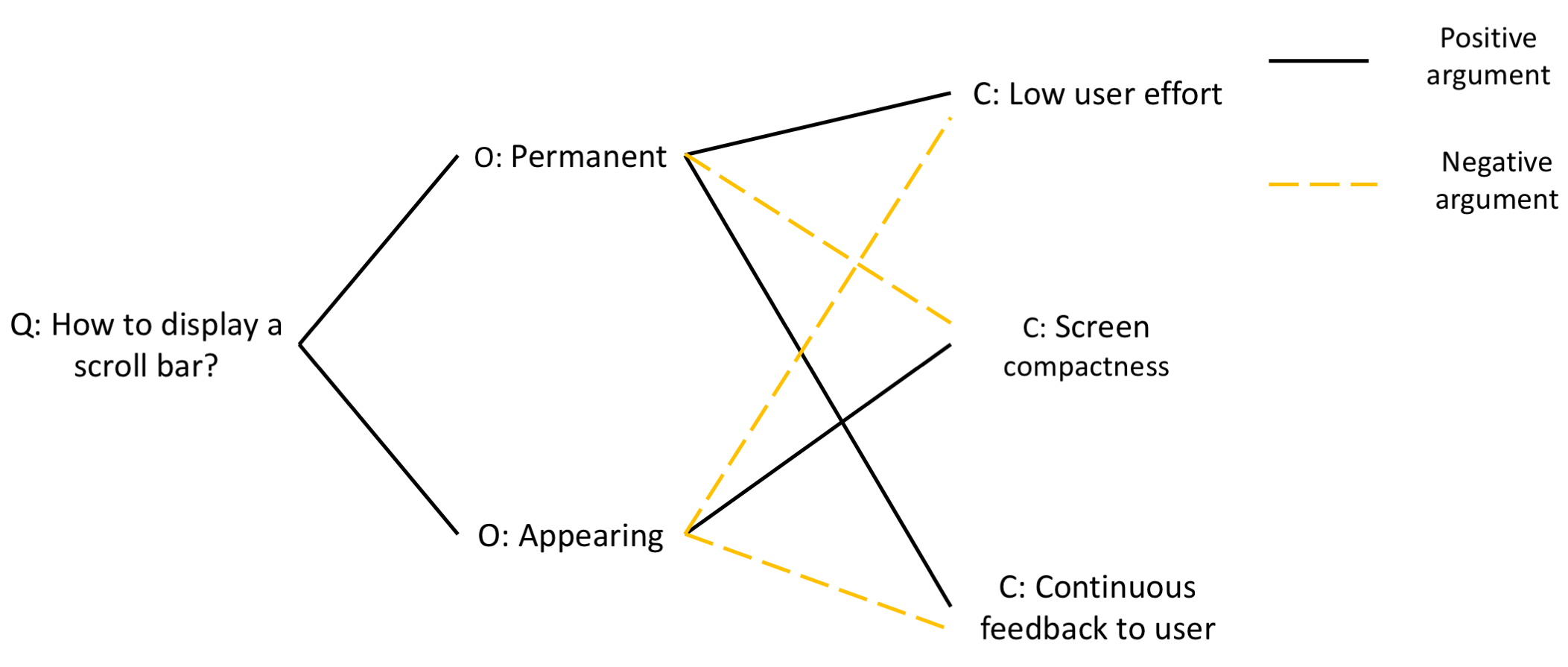}
	\caption{An illustration of the QOC notation. Adapted from \cite{MacLean:1991:QOC:1456153.1456155}.}
	\label{fig:QOC}
	
\end{figure}

One major approach to representing design rationale is design space analysis.  It uses a semi-formal notation called QOC (Questions, Options, and Criteria): \textit{Questions} identify key design issues, \textit{Options} provide possible answers to the Questions, and \textit{Criteria} illustrate the criteria used for assessing and comparing the Options. The QOC notation creates an explicit representation of a structured space of design alternatives and the consideration for choosing among the different choices. It has been widely applied in the design of human-machine interface \citep{Sutcliffe:2018kj, Oulasvirta:2017:CSF:3149825.3131608, MacLean:1991:QOC:1456153.1456155}. Figure \ref{fig:QOC} illustrates the QOC representation of the design space of the scroll bar mechanism in the Xerox Common Lisp (XCL) environment. The design question of interest is how to display the scroll bar. There are two options: the scroll bar appears permanently or only appears when the cursor is moved over the edge of a window. Choosing among various options requires a range of considerations and reasoning over those considerations. For the scroll bar example, the reasoning criteria includes low user effort, screen compactness and continuous feedback to the user. The QOC notation also provides a way to represent an \textit{Assessment} of whether an option does or does not satisfy a criterion. The ``permanent'' option ensures continuous feedback and low user effort at the cost of screen space.

\section{2. THE PRESENT STUDY}

In the present study, we proposed the \textit{option-centric rationale} display for enhancing autonomy transparency, and evaluated its effectiveness via a human-in-the-loop experiment. In the experiment, a human operator uncovered a map for treasures with the help from an intelligent assistant. The intelligent assistant's decision-making rationale are conveyed in the \textit{option-centric rationale} display. We tested the following hypotheses:

First, the \textit{option-centric rationale} display explicitly explores the option space (i.e. all the possible options/actions that an autonomy could take on) and present the rationale why a particular option is the most appropriate at a given context. We expected that the enhanced transparency will lead to higher trust:

\noindent\textbf{H1}: When the \textit{option-centric rationale} display is present, human agents will have higher trust in the autonomous agent.

Second, trust has been recognized as one, if not the most, crucial factor determining the use of automated or autonomous technologies \citep{Hoff:2015fuba}. Trust in automation is defined as the "attitude that an agent will help achieve an individual's goals in situations characterized by uncertainty and vulnerability \citep{lee2004}". The underlying idea is that a human agent’s decision whether or not to use an automated or autonomous technology depends on the extent to which he or she trusts it. When trust is high, the human agent is more likely to use the technology. 
Besides the research focus on trust, several researchers have proposed that the human agent's dependence behavior is inversely related to his or her self-confidence in performing the task by themselves manually without the assistance of the intelligent agent \citep{DeVries2003, Kantowitz:1997,moray2000adaptive}. Therefore, we hypothesized that:

\noindent\textbf{H2}: Trust and self-confidence affect people's acceptance behaviors. Higher trust will lead to higher acceptance and higher self-confidence will reduce acceptance. 

Third, \cite{Parasuraman:1997} categorized different types of automation use and showed disusing a reliable automation harmed task performance. 
In the present study, the intelligent assistant is a knowledge-based agent and reasons using propositional logic. Therefore, we hypothesized that:

\noindent\textbf{H3}: Given a highly capable autonomous agent, higher acceptance leads to higher performance. 

Last, most existing studies on trust in automation, or more recently, trust in autonomy, measured trust once at the end of an experiment via a questionnaire (a "snapshot" of trust).  Only a limited number of studies have viewed trust as a dynamic variable that can strength or decay over time  \citep{Manzey:2012hm, Yang:2017:EEU:2909824.3020230, Lee:1992it, Yang2016}. Prior studies showed that human agents calibrate their trust based on their experience interacting with the automated or autonomous technology \citep{Yang:2017:EEU:2909824.3020230}. With enhanced autonomy transparency, we expected to see a faster trust calibration process: 

\noindent\textbf{H4}: People adjust their trust in the autonomous agent as they gain more experience working with the autonomous agent. In particular, with the \textit{option-centric rationale} display, the rate of adjustment will be faster.







 \section{3. METHOD}
 This research complied with the American Psychological Association code of ethics and was approved by the Institutional Review Board at the University of Michigan. 
 
\subsection{3.1 Participants}
Thirty-four participants (Age: Mean = 21.17 years, \textit{SD} = 1.66 years) took part in the experiment. All participants had normal or corrected-to-normal sight and hearing. Participants were compensated with \$5 upon completion of the experiment. In addition, there was a chance to obtain an additional bonus of 1 to 20 dollars based on their performance.

\subsection{3.2 Simulation testbed}

\begin{figure} [h]
	\centering
	\subfloat[\label{fig:wumpus world}]{\includegraphics[width=0.35\linewidth]{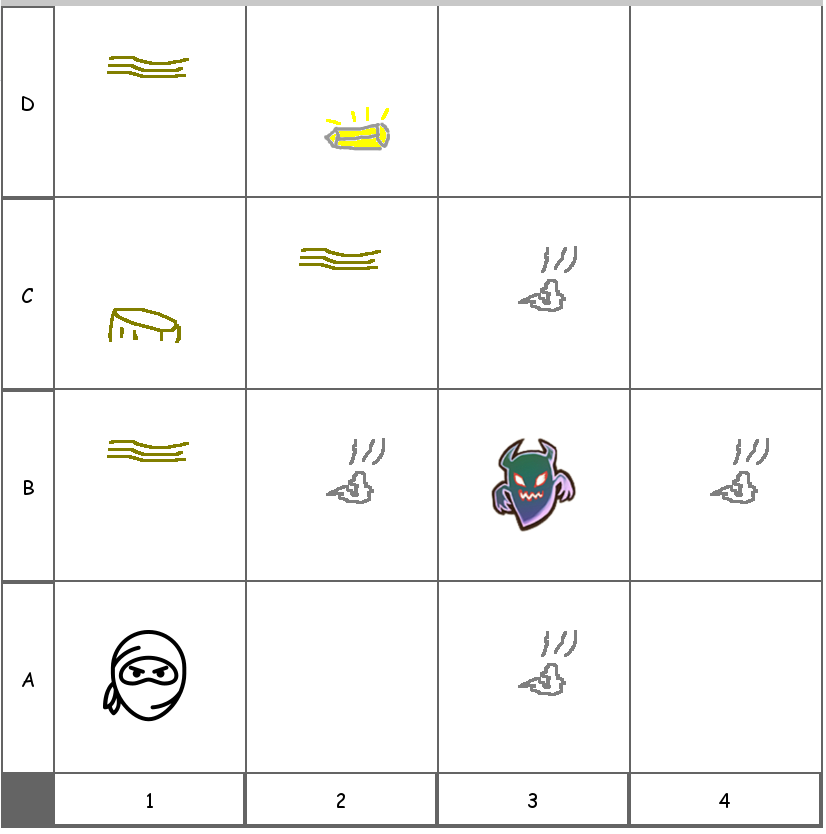}}
	\hspace{0.5cm}
	\subfloat[\label{fig:action}]{\includegraphics[width=0.5\linewidth]{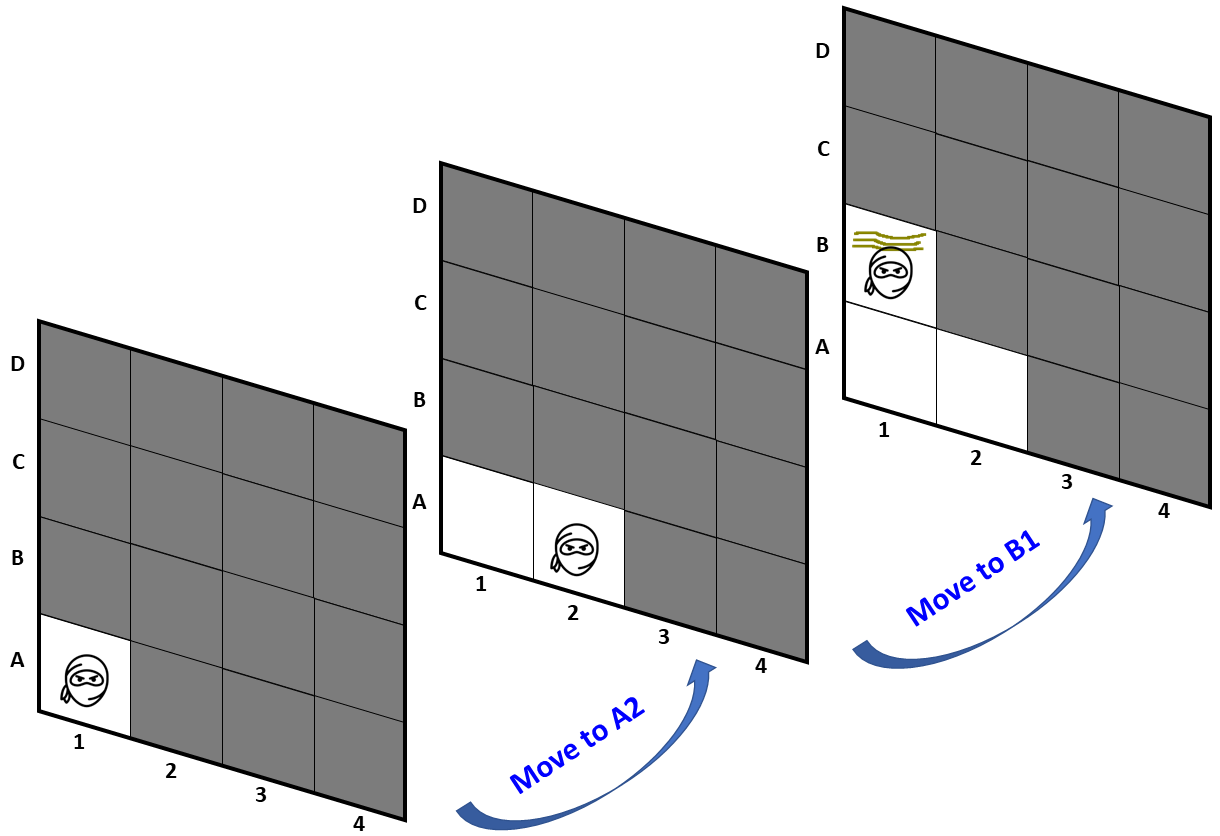}}
	\caption{(a) An example map in Treasure Hunter. Each square is denoted by the row number (from 1 to 4) and the column number (from A to D). (b) First two steps of a hunter moving in the map.}
	\label{fig:wumpus}
\end{figure}

We developed an experimental testbed -- \textit{Treasure Hunter}, adapted from the Wumpus world game \citep{Russell:2010}. In the game, the participant acts as a hunter to find the gold bar in the map with the help of an intelligent assistant (Figures~\ref{fig:wumpus world} \& \ref{fig:action}). Each step, the hunter can move to an unvisited location which is connected to the visited locations. Figure~\ref{fig:action} shows that the hunter moves from A1 to A2 and then to B1. On the way to the treasure, the hunter might fall into a pit (shown in C1 in Figure~\ref{fig:wumpus world}) or encounter a wumpus (shown in B3 in Figure~\ref{fig:wumpus world}). The hunter gathers information about his or her surroundings by a set of sensors. The sensors will report a stench when the wumpus is in an adjacent location (shown as B2, A3, C3, B4 in Figure~\ref{fig:wumpus world}) and a breeze when a pit is in an adjacent location (shown as B1, C2, D1 in Figure~\ref{fig:wumpus world}).  There is one and only one gold bar/wumpus in a map. However, there might be one or multiple pits in a map. Each element - a pit, a wumpus, or a gold bar - occupies a unique location on the map.

Table~\ref{tab:score_conseq} shows the scores and consequences for different events. If the hunter finds the gold bar, s/he will receive 500 points and the game will end. If the hunter encounters the wumpus, s/he will lose 1000 points and the game will end. If the hunter falls into a pit, s/he will lose 100 points but can still continue the game. The hunter will only fall into a pit at the first time he encounters it. The hunter will get a 10-point penalty for uncovering every new location. 

\begin{table}[H]
\caption{Scores and consequences for different events}
\label{tab:score_conseq}
\begin{tabular}{l|l|l}
\hline
Event                     & Score & Consequence                                \\ \hline
Find the gold bar                 & +500  & Map ends                                   \\ \hline
Discover one new location & -10   & Continue                                   \\ \hline
Fall into a pit             & -100  & Continue, no more points lost when revisit \\ \hline
Meet wumups               & -1000 & Map ends                                   \\ \hline
\end{tabular}
\end{table}

\begin{figure}
    \centering
    \includegraphics[width=0.32\linewidth]{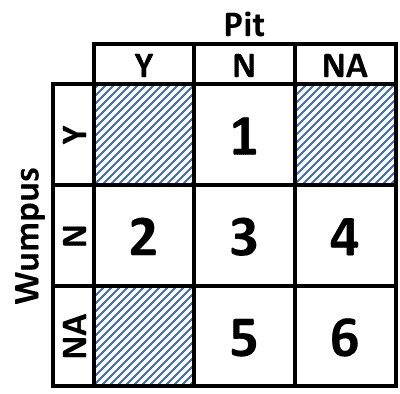}
    \hspace{8mm}
    \begin{tabular}[b]{c|c|c}
    \hline
    ID & P(W, P, G, N)                                     & Expected Score   \\\hline\hline
    1  & $(1, 0, 0, 0)$                                    & $-1000$          \\\hline
    2  & $(0, 1, 0, 0)$                                    & $-100$           \\\hline
    3  & $(0, 0, 0.5, 0.5)$                                & $250$            \\\hline
    4  & $(0, \frac{1}{3}, \frac{1}{3}, \frac{1}{3})$      & $133.33$            \\\hline
    5  & $(\frac{1}{3}, 0, \frac{1}{3}, \frac{1}{3})$      & $-166.67$        \\\hline
    6  & $(\frac{1}{4}, \frac{1}{4}, \frac{1}{4}, \frac{1}{4})$                        & $-150$           \\\hline
    \end{tabular}
    \captionlistentry[table]{A table beside a figure}\label{tab:case_condition}
    \captionsetup{labelformat=andtable}
    \caption{Six potential cases with the corresponding probabilities/expected scores based on the reasoning of pit and wumpus conditions from the intelligent assistant. Y: there is a pit/wumpus. N: there is no pit/wumpus. NA: it is not sure to have a pit/wumpus. Shaded squares are the impossible cases because the pit and wumpus cannot co-exist in one location. W: encounter a wumpus; P: fall into a pit; G: find a gold bar; N: nothing happens.}
    \label{fig:Different_cases}
\end{figure}

An intelligent assistant helps the participant by recommending where to go. The intelligent assistant is a knowledge-based agent and reasons using propositional logic \citep{Russell:2010}. Propositional logic is a mathematical model that reasons about the truth or falsehood of logical statements. By using logical inference, the agent will give the values of four logical statements for a given location (e.g. location $D2$): (1) there is a pit at this location, denoted as $P_{D,2}$; (2) there is no pit at this location, denoted as $\neg P_{D,2}$; (3) there is a wumpus at this location, denoted as $W_{D,2}$; (4) there is no wumpus at this location, denoted as $\neg W_{D,2}$. Based on the value of these 4 logical statements, we can categorize the location into one of the six different conditions shown in Figure~\ref{fig:Different_cases}: Y represents there is a pit/wumpus at this location (value of the first/third logical statements is true); N represents there is no pit/wumpus at this location (value of the second/fourth logical statement is true); NA represents the agent is not sure about the existence of pit/wumpus at this location (values of all the four statements are false). The shaded squares in Figure~\ref{fig:Different_cases} are the impossible cases because the pit and wumpus cannot co-exist in one location. For each case in Figure~\ref{fig:Different_cases}, the agent will assign probabilities of encountering a wumpus, falling into a pit, finding a gold bar or nothing happens as well as the corresponding expected scores if the hunter moves to that location as shown in Table~\ref{tab:case_condition}. The agent will randomly select one of the potential next locations with the highest expected score as the recommendation. 

Every step during the experiment, the participant will first receive the suggestion from the intelligent assistant, and then make a decision, i.e. select the target location that s/he wants to go next. After the participant makes a decision, the hunter will move to the next location and the intelligent assistant will update its knowledge based on the sensory feedback (breeze and stench at the new location). The intelligent agent is a level 3/4 automation as defined by \citet{Sheridan1978} wherein the autonomy narrows down the action selections and suggests one alternative to the human operator. Similar experimental paradigms have been used to examine trust in and dependence on automation \citep{Manzey:2012hm,Neyedli2011,Wang2009Trust,McBride:2011ix}.


\begin{figure} [h]
	\centering
	\includegraphics[width=\linewidth]{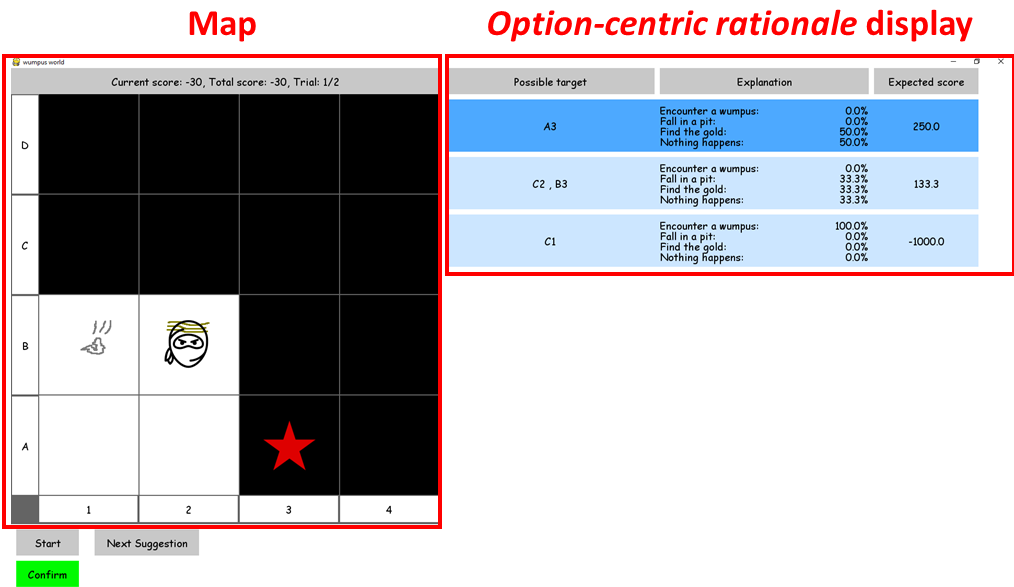}
	\caption{Testbed with \textit{option-centric rationale} display.}
	\label{fig:exp_testbed}
\end{figure}


 Figure~\ref{fig:exp_testbed} shows the \textit{option-centric rationale} display proposed in this study. The \textit{option-centric rationale} display details all the available next locations and the criteria for choosing a particular location, and highlights the final recommendation using a red star. The criteria for recommending a particular location depends on whether the human-autonomy team will find the gold bar, fall into a pit, encounter a wumpus, or uncover a new location (without finding a gold bar, falling into a pit or encountering a wumpus). The display also shows the possibility of each criterion and the corresponding expected score. The display will group the next locations based on the criteria, i.e. if two locations have the same probabilities of each criterion, the display will list them in the same row. The locations are sorted from the highest expected score to the lowest. The final recommendation is one of the locations with the highest expected score. Note that the available next locations, the possibility of each criterion and the expected scores are all computed by the intelligent assistant.

  

\subsection{3.3 Experimental design}
The experiment used a within-subjects design. The independent variable in the experiment was the presence/absence of the \textit{option-centric rationale} display. The order of the two conditions was counterbalanced to eliminate potential order effects. In each condition, participants played the game on 5 different maps. In the absence of the \textit{option-centric rationale} display condition, the participant only saw a red star that indicated the recommendation by the intelligent assistant.

\subsection{3.4 Measures}
We measured three groups of dependent variables: subjective responses, behavioral responses and performance. After completing each map, participants were asked to report their trust in the intelligent assistant and their self-confidence to accomplish the task without the intelligent assistant using two 9-point Likert scales: (1) How much do you trust the intelligent assistant? (2) How confident are you in completing tasks without the intelligent assistant? We calculated the recommendation acceptance as the rate that the participant followed the recommendations given by the intelligent assistant. 
Participants' scores for each map were recorded as well.

\subsection{3.5 Map selection}
\begin{figure}[h]
\centering
\subfloat[Training map 1]{\includegraphics[width = 0.2\linewidth]{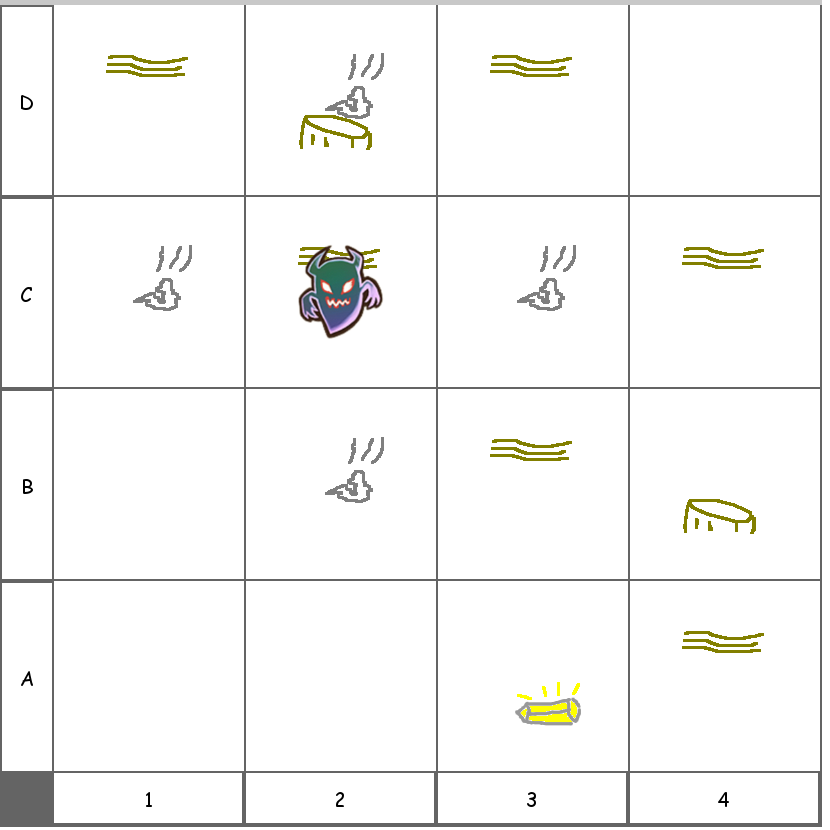}}
\subfloat[Training map 2]{\includegraphics[width = 0.2\linewidth]{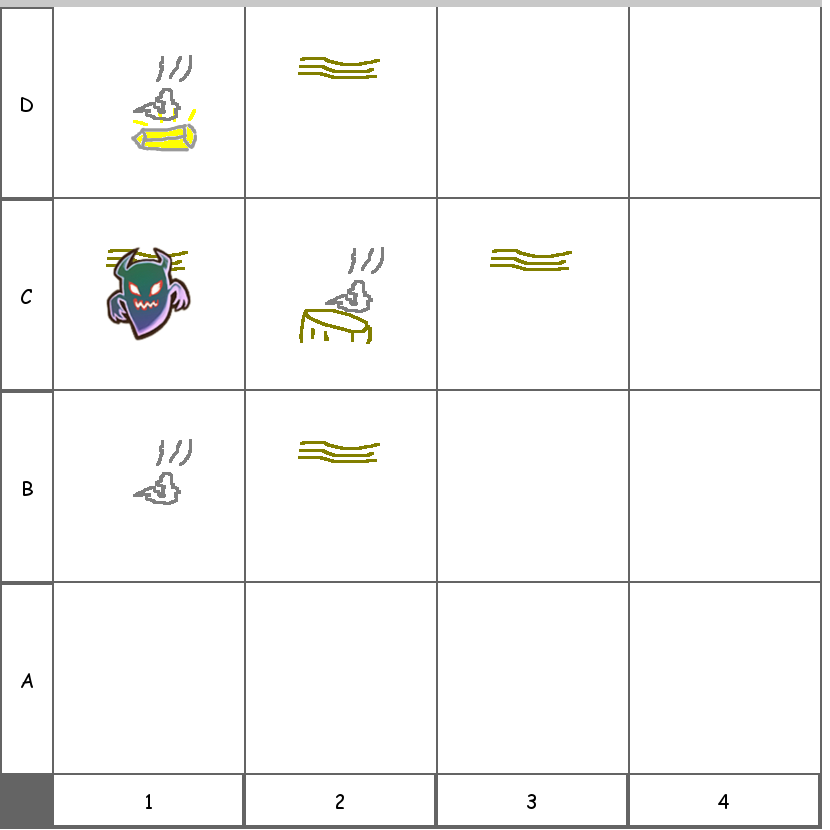}}
\subfloat[Training map 3]{\includegraphics[width = 0.2\linewidth]{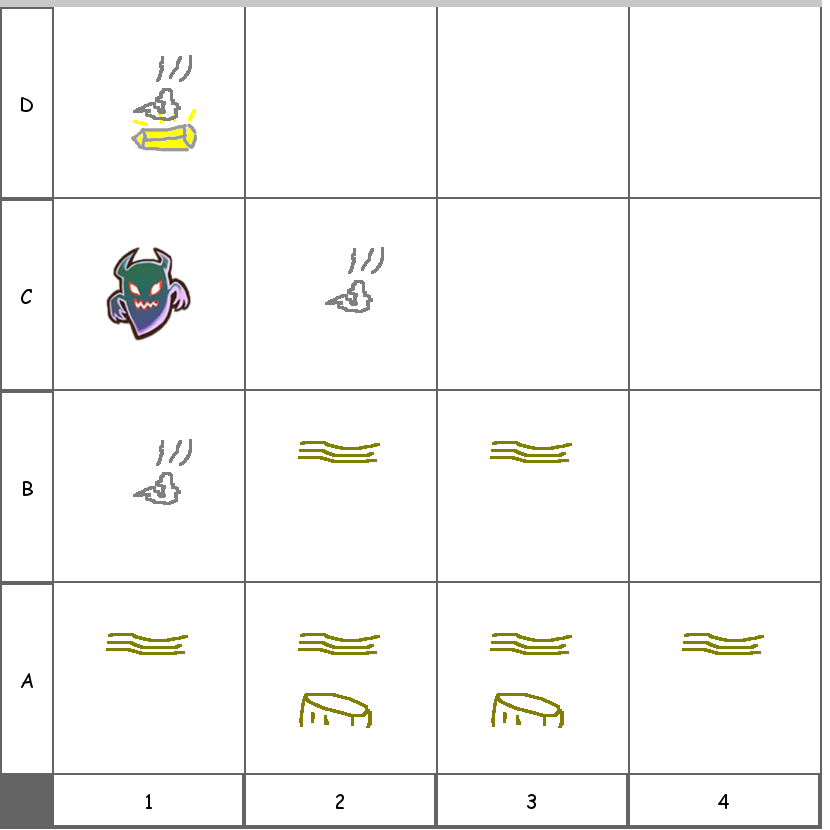}}
\subfloat[Training map 4]{\includegraphics[width = 0.2\linewidth]{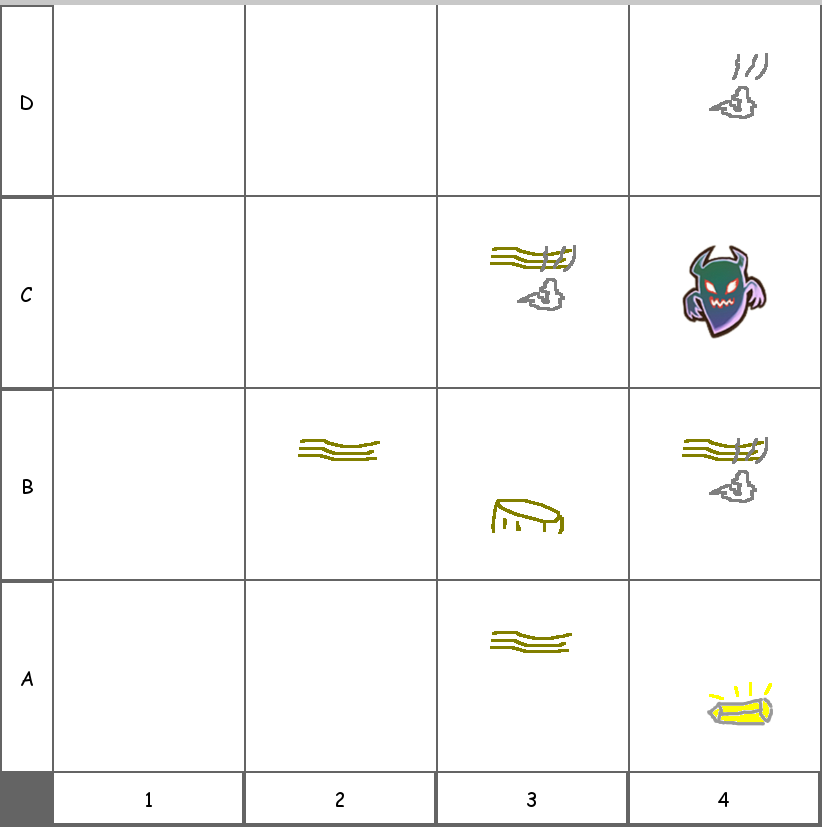}}
\subfloat[Training map 5]{\includegraphics[width = 0.2\linewidth]{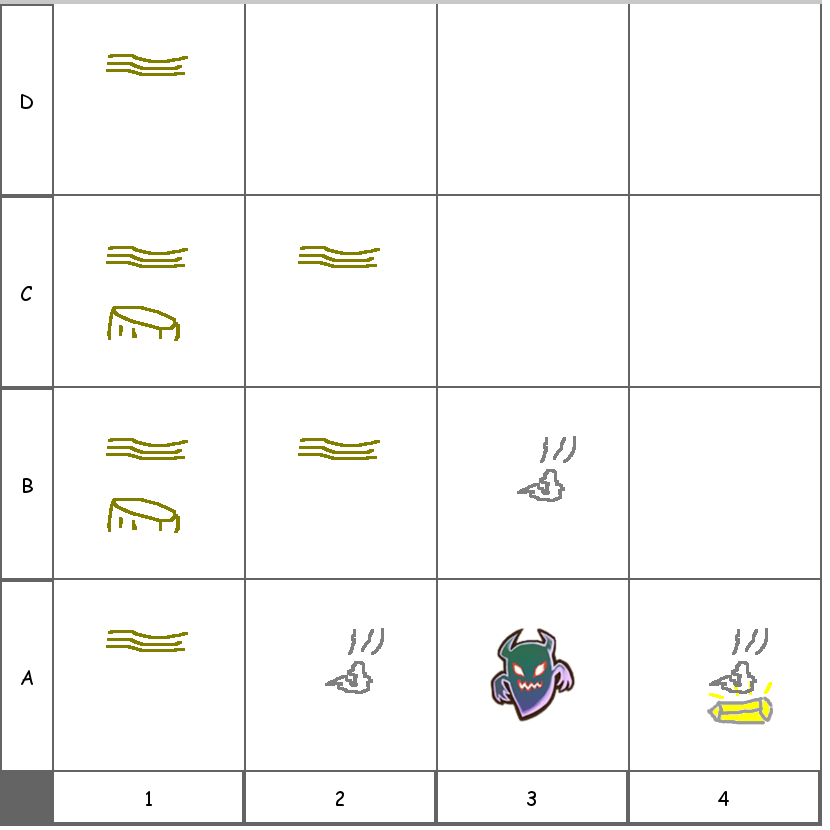}}

\centering
\subfloat[Testing map 1]{\includegraphics[width = 0.2\linewidth]{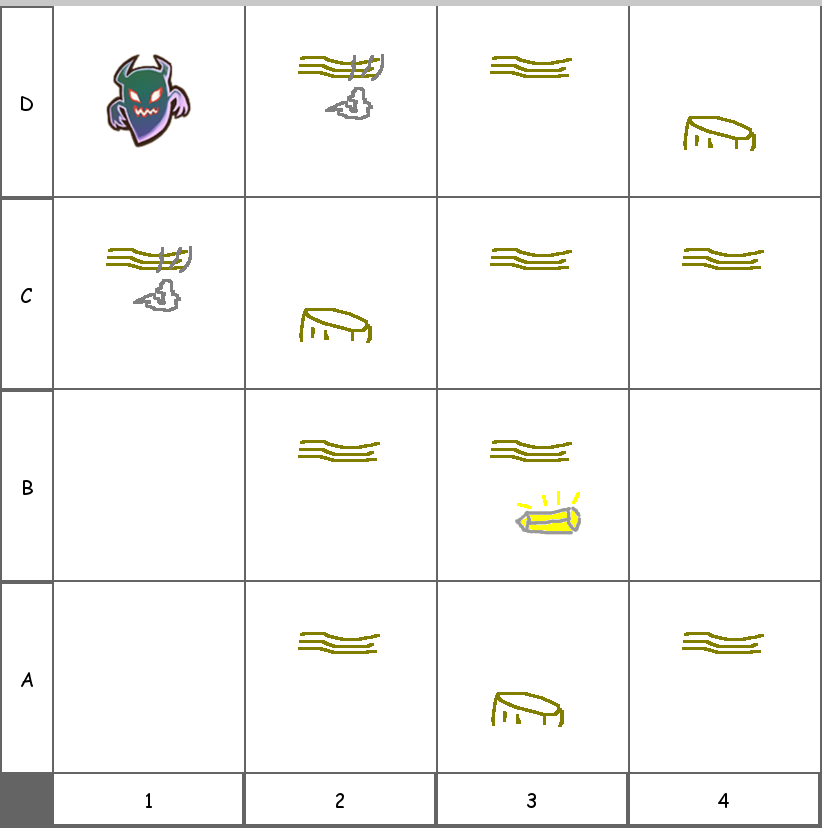}}
\subfloat[Testing map 2]{\includegraphics[width = 0.2\linewidth]{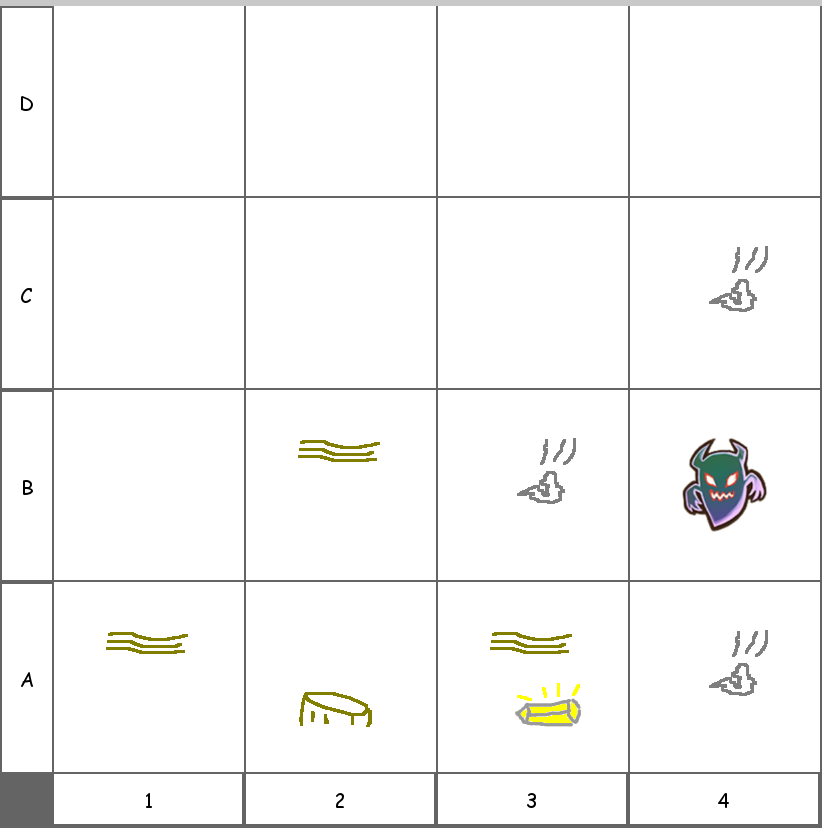}}
\subfloat[Testing map 3]{\includegraphics[width = 0.2\linewidth]{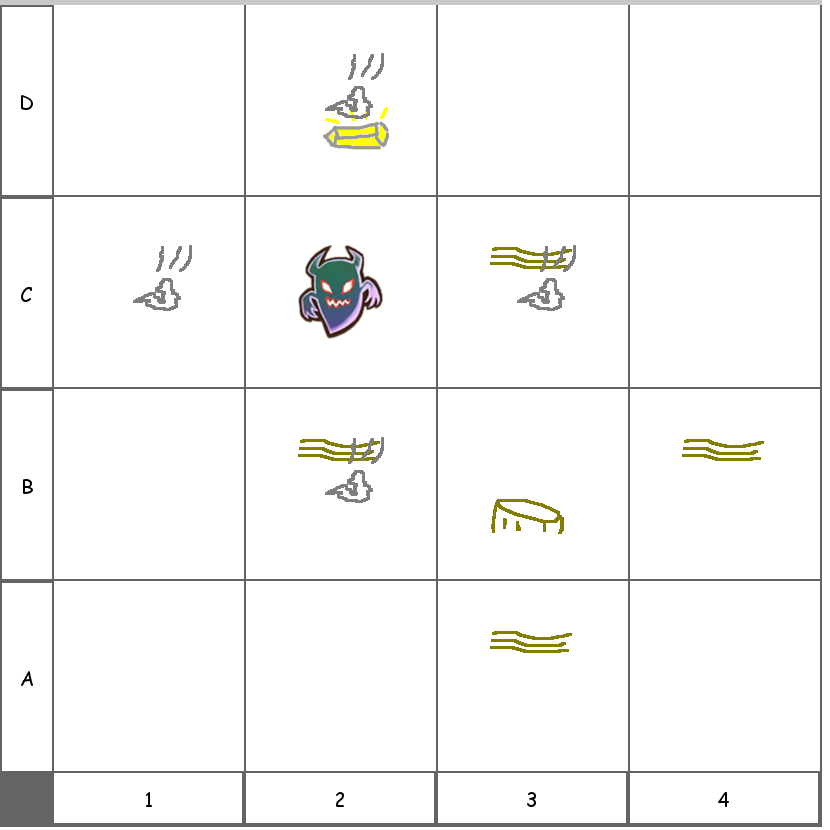}}
\subfloat[Testing map 4]{\includegraphics[width = 0.2\linewidth]{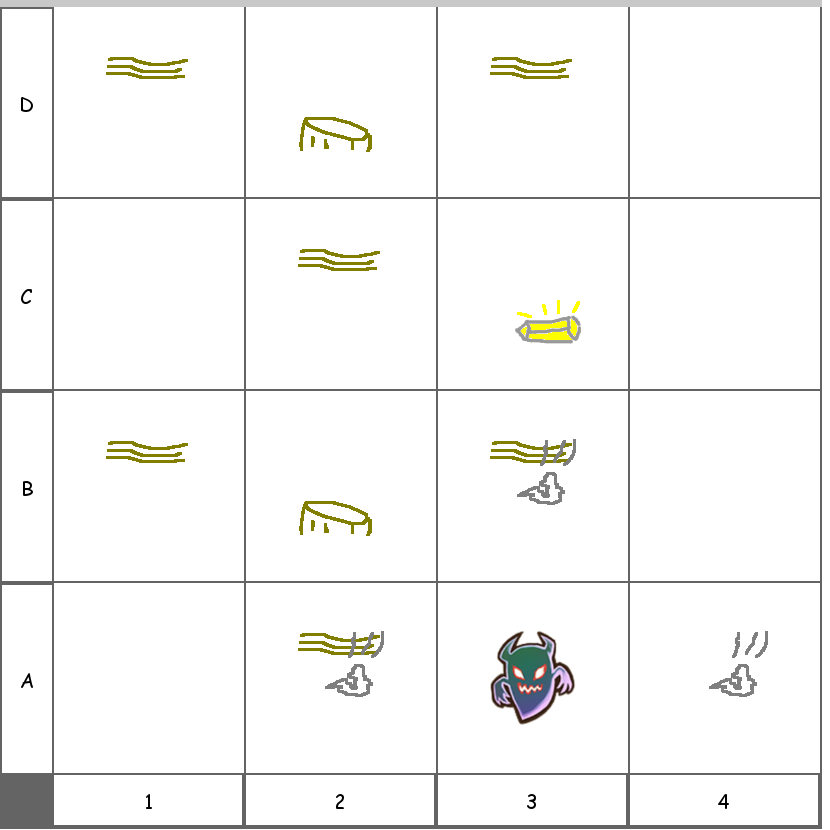}}
\subfloat[Testing map 5]{\includegraphics[width = 0.2\linewidth]{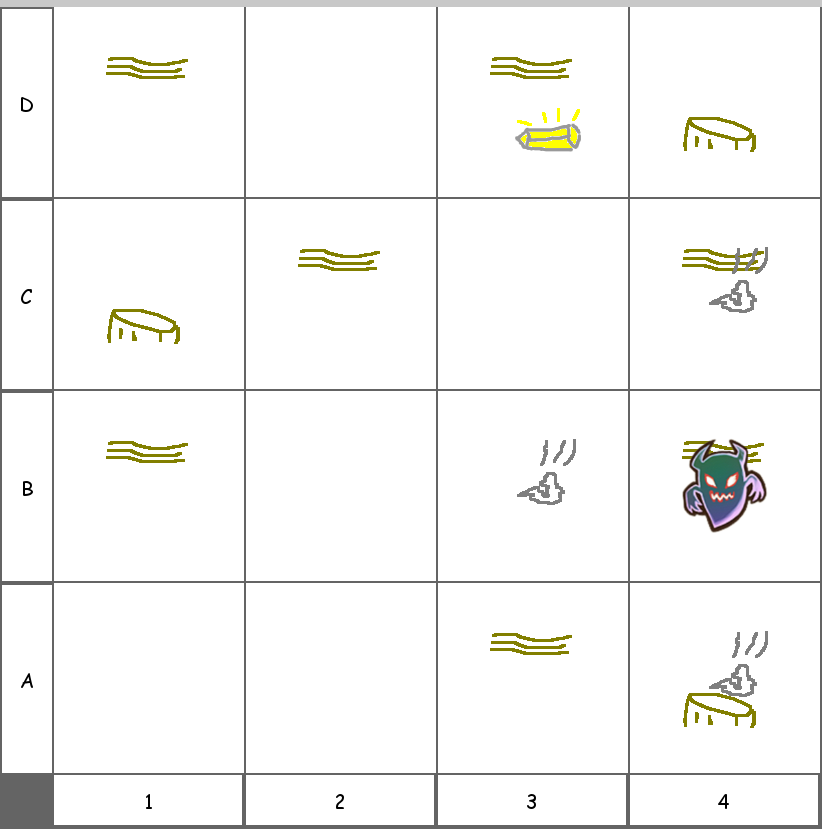}}

\centering
\subfloat[Testing map 6]{\includegraphics[width = 0.2\linewidth]{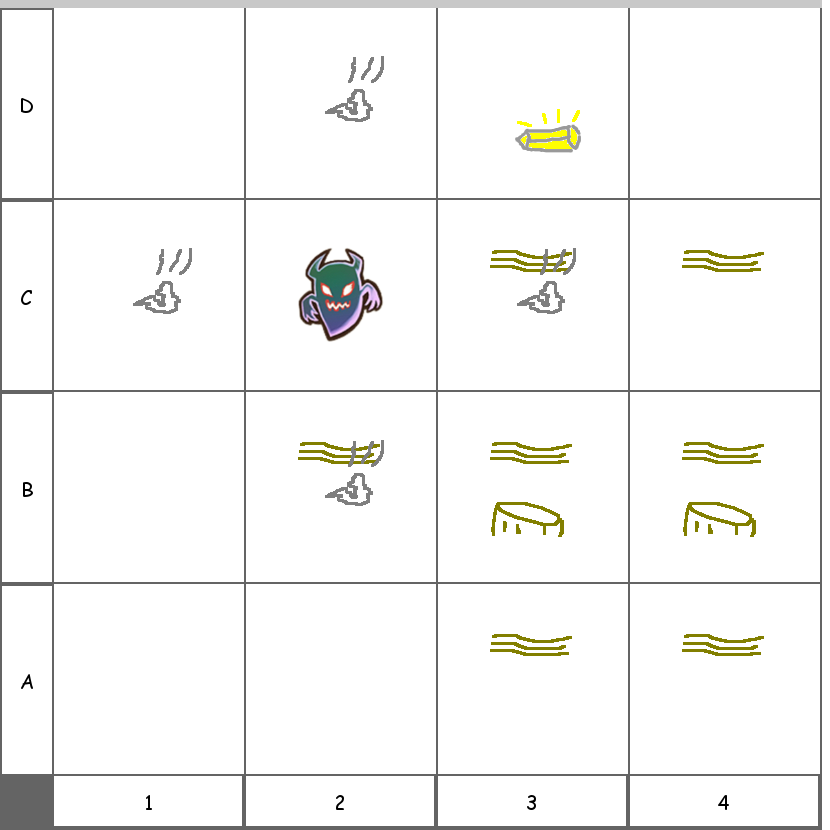}}
\subfloat[Testing map 7]{\includegraphics[width = 0.2\linewidth]{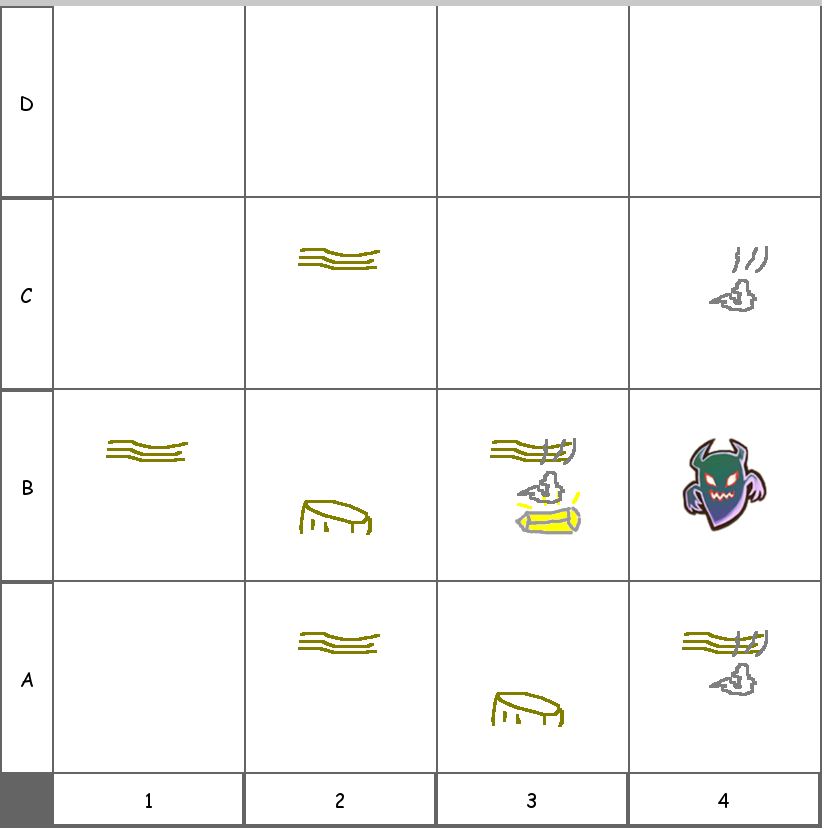}}
\subfloat[Testing map 8]{\includegraphics[width = 0.2\linewidth]{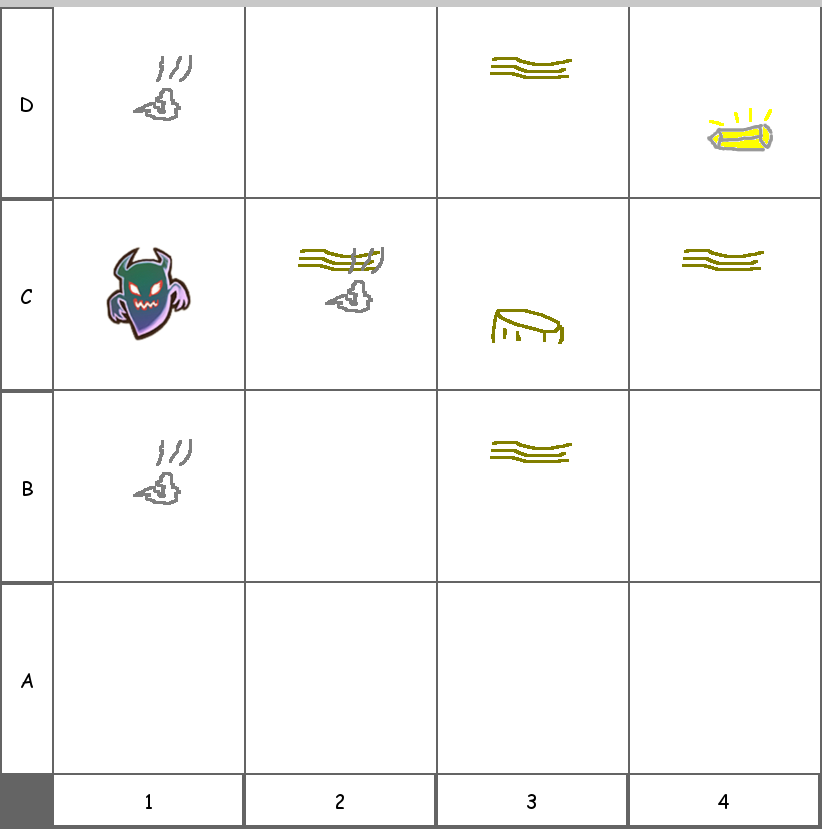}}
\subfloat[Testing map 9]{\includegraphics[width = 0.2\linewidth]{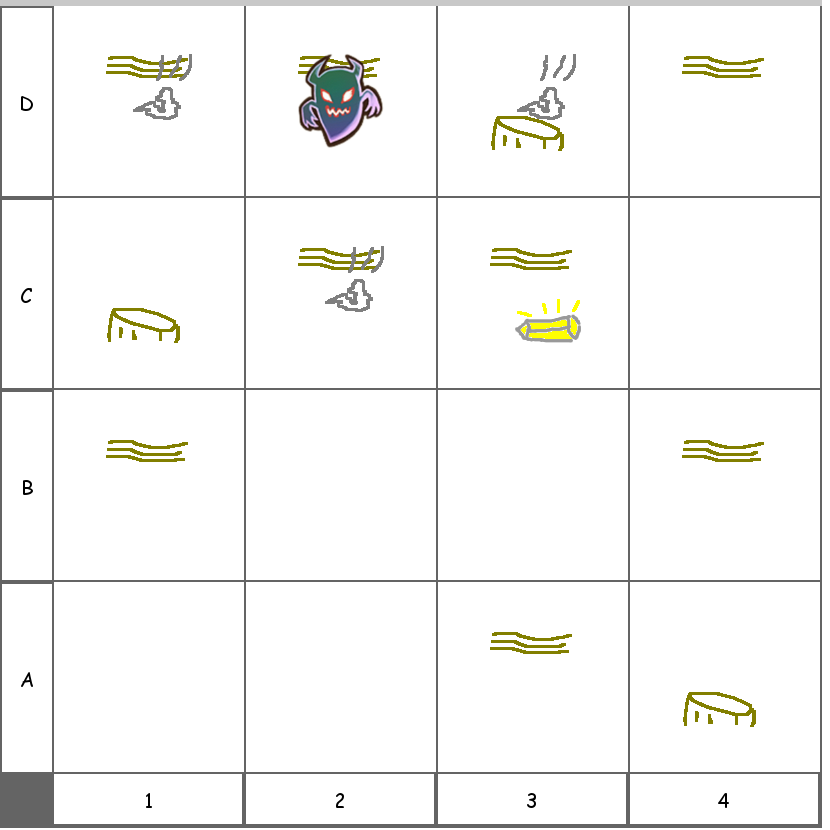}}
\subfloat[Testing map 10]{\includegraphics[width = 0.2\linewidth]{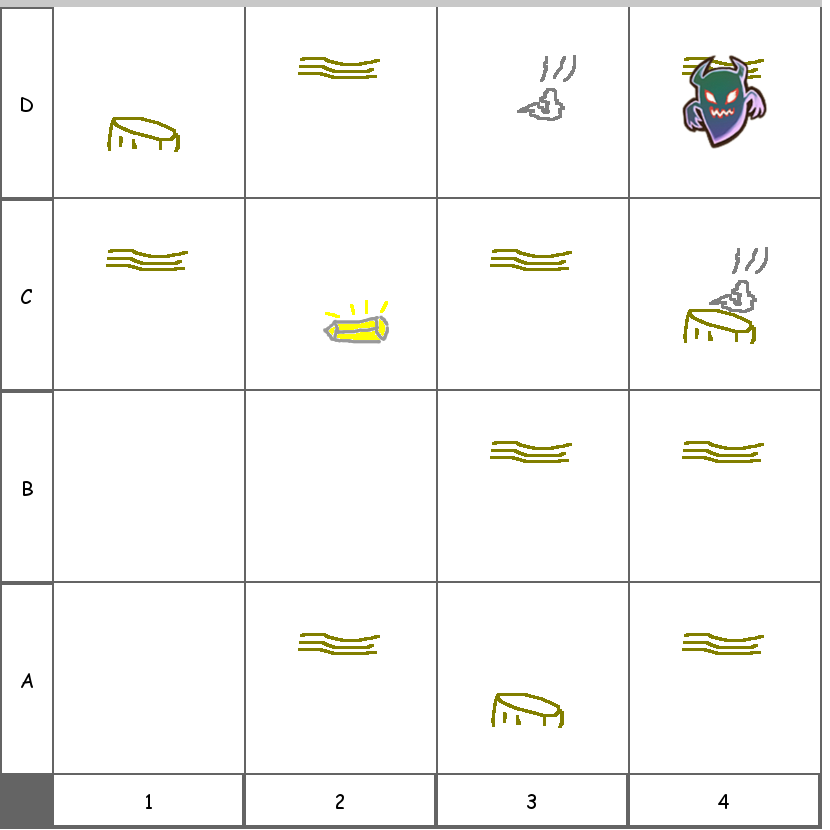}}

\caption{Selected maps for training and testing. First row: fixed order training map. Second and third row: testing map, order is randomly determined for each participant.}
\label{fig:selected_maps}
\end{figure}

 In order to eliminate the inherent randomness of the task, we carefully selected the maps used in the experiment (Figure \ref{fig:selected_maps}). First, we randomly generated 100 maps and ran the game only with the intelligent assistant 20 times for each map (i.e, always accepted the recommendations from the intelligent assistant). We ranked the maps based on the standard deviation of the scores for each map from the lowest to the highest. Second, we selected 10 maps which fulfilled three criteria: (1) Each map had a low standard deviation of the scores; (2) In each map, the gold bar was not just next to the start location; (3) The locations of the gold bar in the 10 maps should be balanced across the maps instead of concentrating in one part of the maps (e.g. upper right corner of the map). For each participant, the order of the 10 maps in the experiment are randomly determined. The second row in Table~\ref{tab:map_score} shows the mean and standard error of the intelligent assistant's score of the 10 selected maps.
 
 We also developed 5 maps for the training session. Out of the 5 training maps, there are two maps with a pit next to the start location and three maps with low standard deviation of scores. The 5 training maps were presented according to the following order: The first was similar to the maps participants experience in the real test. Participants 
 practiced on this map without the help of the intelligent assistant. The aim was to help participants get familiar with the game. From the second map onward, participants played the game with the help of the intelligent assistant. The second and fourth practice maps were similar to the maps participants experienced in the real test. The third and the fifth maps contained a pit next to the start location. The reason for selecting the two maps (i.e. the third and the fifth map) was to help participants fully understand the stochastic nature of the game. For example, in the fifth training map (Figure~\ref{fig:selected_maps}), a breeze was detected by the sensor at the start location and the two adjacent locations (i.e. B1 and A2) have the same probability of having a pit.

\subsection{3.6 Procedure}
All participants provided informed consent and filled in a demographics survey. After that, participants received a practice session. Participants played the game first without the intelligent assistant, and practiced on another four maps with the intelligent assistant, and with or without the \textit{option-centric rationale} display. In the experiment, participants played the game with 5 maps in each condition. After each map, participants were asked to report their trust in the intelligent assistant and their confidence in accomplishing the game without the help of the intelligent assistant. Participants' acceptance behaviors and task performance were recorded automatically by the testbed.



\section{4. RESULTS}

Data from 4 participants were discarded due to malfunction of the testbed. Data from 2 participants were discarded as their task performance were considered as outliers based on the two-sided Dixon's Q test \citep{dixon1953processing}. All hypotheses were tested using data from the remaining 28 participants (Mean age = 21.25 years, SD = 1.72 years). To analyze the data, we performed structural equation modeling (SEM) using SPSS AMOS. SEM is a multivariate statistical method widely used in behavioral science \citep{Bollen2011, Goldberger1972}. It uses a path diagram to specify the model, which indicates the relationships between the variables. An arrow represents an effect of a variable on another. In SEM, multiple regressions are simultaneously estimated to indicate the strength of each relationship (arrow).  Measurement of fit in SEM provides information about how well the model fits the data. A good-fit model is required before interpreting the causal paths of the structural model.


\subsection{4.1 Relationships between the \textit{option-centric rationale} display, trust, self-confidence, recommendation acceptance and performance}

\begin{figure}[h]
\hspace{20mm}
    \centering
    \includegraphics[width=1\textwidth]{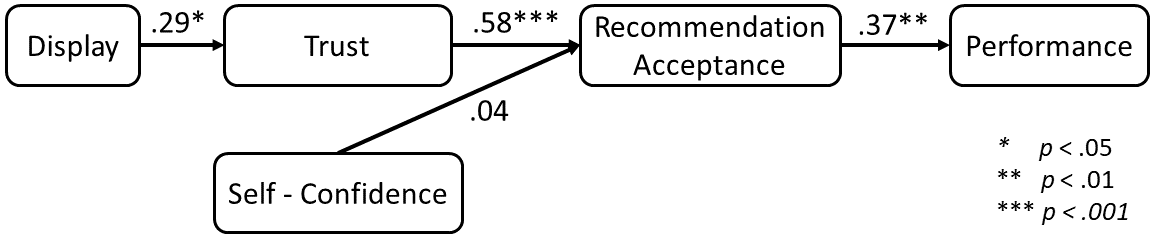}
    \caption{The structural equation model indicates the relationship between the display, trust, self-confidence, acceptance and task performance.}
    \label{fig:SEM}
    \vspace{-5mm}
\end{figure}

To test hypotheses 1, 2 and 3, we constructed and tested a model as specified in Figure \ref{fig:SEM}. Based on prior research \citep{steiger2007understanding,hooper2008structural, hu1999cutoff}, multiple model fit indices were used to check the model fit. Our model demonstrated a good fit, with $\chi^2(6) = 3.776, p = .71$, $\text{RMSEA} = 0.00$, $\text{CFI} = 1.00$, $\text{SRMR} = .058$ .

The model is shown in Figure ~\ref{fig:SEM}, with annotated standardized weights and significance (noted with asterisks). Standardized weights can be directly compared to assess the strength of each factor’s impact on the subsequent factor. In our study, the chains connecting display type, trust, recommendation acceptance, and task performance were significant: the \textit{option-centric rationale} display increased participants' trust, which significantly affected their recommendation acceptance, and in turn impacted the task performance. Yet, participants' self-confidence did not affect their recommendation acceptance.

\subsection{4.2 Trust adjustment overtime}
To examine how trust changes overtime, we conducted a linear mixed effects analysis, with the presence/absence of the \textit{option-centric rationale} display and trial number as fixed effects and intercepts for subjects as random effects. Results are reported as significant for $\alpha < .05$. 

\begin{figure}[h]
    \vspace{-5mm}
    \centering
    \includegraphics[width=1\textwidth]{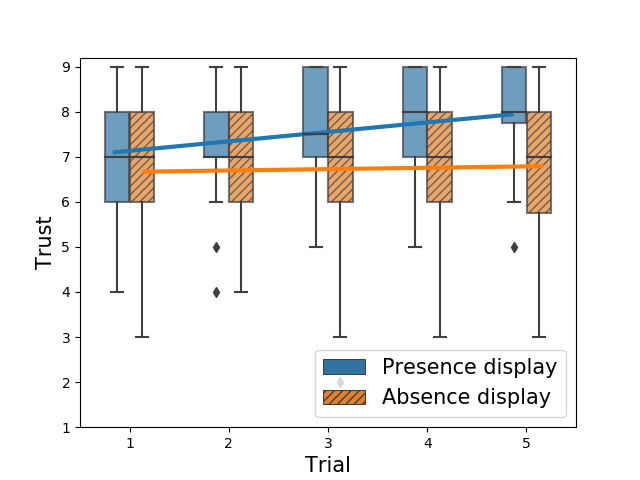}
    \caption{The change of trust over time.}
    \label{fig:trust_time}
\end{figure}

There was a significant interaction effect between display and time, $F(1,249) = 5.675, p = .018$ (See Figure~\ref{fig:trust_time}). Specifically, when participants were provided with the \textit{option-centric rationale} display, their trust in the autonomous agent increased as they gained more experiences, $F(1,111) = 43.714, p < .001$. When the display was not present, however, their trust did not change significantly, $F(1,111) = 0.549, p = 0.460$.

\section{5. DISCUSSION}



 



In this study, we proposed an \textit{option-centric rationale} display. The display explicitly conveys the intelligent assistant's decision-making rationale by detailing all the available next locations and the criteria for recommending a particular location.



Our results indicate that with the \textit{option-centric rationale} display, human operators trusted the intelligent assistant more. In addition, consistent with findings from previous studies \citep{Koo:2014eg,Koo:2016hx,Forster:2017cg,Beller2013}, we also found that trust significantly affected people's acceptance behaviors. As human operators' trust in  the autonomous agent increased, they accepted more recommendations provided by the intelligent assistant. Meanwhile, human operators' self-confidence did not influence their recommendation acceptance, which was in line with previous research \citep{Lee:1994hk, moray2000adaptive}.


We found that higher recommendation acceptance led to higher task performance. We argued that this positive relationship hinged on the capability of the intelligent assistant, which was near-optimal in our study. 
Table~\ref{tab:map_score} details the optimal score that an omniscient agent could obtain and the score that the knowledge based intelligent assistant used in the present study could obtain. The optimal score was calculated assuming that the intelligent assistant was omniscient (i.e., the map was known to the intelligent assistant). The intelligent assistant's score was calculated by having the autonomous agent play the treasure hunter game by itself for 20 times. 
The intelligent assistant's performance was close to the optimal score. The ratio between the intelligent assistant's score and the optimal score was on average $91.1\%$. 


\begin{table}[t]
\centering
\caption{Mean and Standard Error (SE) values of the Intelligent Assistant's Score and the Optimal Score for Each Test Map}
\label{tab:map_score}
\begin{adjustbox}{width=\linewidth}
\begin{tabular}{ccccccccccc}
\hline
Test Map ID              & 1           & 2             & 3               & 4           & 5             & 6             & 7               & 8               & 9               & 10              \\ \hline
Intelligent Assistant's Score & $450 \pm 0$ & $381 \pm 2.4$ & $430.5 \pm 0.5$ & $440 \pm 0$ & $415 \pm 1.1$ & $421 \pm 0.7$ & $363.5 \pm 3.5$ & $386.5 \pm 3.4$ & $428.5 \pm 1.7$ & $447.5 \pm 1.9$ \\ \hline
Optimal Score      & 470         & 460           & 460             & 460         & 450           & 450           & 450             & 440             & 460             & 470             \\ \hline
Ratio (\%)& 95.7&82.8&93.6&95.7&92.2&93.6&80.8&87.8&93.2&95.2\\\hline
\end{tabular}
\end{adjustbox}
\end{table}

Prior research showed that as people gained more experience interacting with the autonomous agent, their trust in the agent will evolve and eventually stabilize if enough interaction is warranted \citep{Yang:2017:EEU:2909824.3020230}. Results from this study supported prior results. More importantly, human operators adjusted their trust faster when the \textit{option-centric rationale} display was presented. This implies that human operators would require less amount of interaction with the autonomous agent to calibrate their trust before trust reaching a steady state. This benefit is also attributable to the enhanced transparency provided by the \textit{option-centric rationale} display. The list of all available options prevents severe drop of trust when optimal performance was not achieved simply because of the inherent randomness in the game. 


Although we only tested the \textit{option-centric rationale} display on a simulated game with a small action space, the display can be applied to other decision-making agents with a larger action space, for instance, an epsilon-greedy agent 
with finite (i.e., countable) action space. The epsilon-greedy agent balances exploration and exploitation by choosing the optimal action some times and the exploratory action other times. The exploratory action is not the optimal action at a particular step. However, by further exploring the environment, the agent can obtain higher rewards in the subsequent steps and higher accumulative rewards. 
The \textit{option-centric rationale} display can list all possible actions with the expected reward, and the number of times the optimal/exploratory action has been taken to indicate the necessity of exploring the environment. For a large action space, the \textit{option-centric rationale} display can present a subspace of the action space that contains the optimal and near optimal actions by listing the actions with top expected scores. The other (far from optimal) actions can be displayed if requested. Further research is needed to determine the size of subspace to be displayed.

\section{6. CONCLUSION}
The advance in artificial intelligence and machine learning empowers a new generation of autonomous systems. However, human agents increasingly have difficulty deciphering autonomy-generated solutions and therefore perceive autonomy as a mysterious black box. The lack of transparency contributes to the lack of trust in autonomy and sub-optimal team performance \citep{Chen:2014cmba,Endsley:2017,Lyons:2014,deVisser:2018ju, Yang:2017:EEU:2909824.3020230, Lyons:2016co}. In this study, we proposed an \textit{option-centric rationale} display for enhancing autonomy transparency. The \textit{option-centric rationale} display details all the potential actions and the criteria for choosing a particular action, and highlights the final recommendation. The results indicate that the presence/absence of the display significantly affected people's trust in the autonomous agent and human operators' trust increased faster when the display was provided. The results should be reviewed in light of several limitations. First, the intelligent assistant used in the present study was highly capable. 
However, in the real world, an intelligent assistant could be less capable in situations of high uncertainty and ambiguity. Further research with less capable autonomous agents is needed to validate the generalization of the display. Second, the action space in the simulated game was limited. We discussed the application of the \textit{option-centric rationale} display on domains with larger action space. Further research is needed to examine the proposed solutions.

\bibliography{main}

\newpage
\section{Biographies}
\textbf{Ruikun Luo} is a Ph.D. candidate at the Robotics Institute, University of Michigan, Ann Arbor. Prior to joining the University of Michigan, he obtained a M.S. in Mechanical Engineering from Carnegie Mellon University in 2014 and a B.S. in Mechanical Engineering and Automation from Tsinghua University, China in 2012. \\


\textbf{Na Du} is a Ph.D. pre-candidate in the Department of Industrial \& Operations Engineering at the University of Michigan. Her research interest includes trust in automation and design of decision aids on trust calibration. Prior to joining the University of Michigan, she completed a B.S. in Psychology in Zhejiang University, China in 2016.\\



\textbf{X. Jessie Yang} is an Assistant Professor in the Department of Industrial and Operations Engineering and an affiliated faculty at the Robotics Institute, University of Michigan, Ann Arbor. She obtained a PhD in Mechanical and Aerospace Engineering (Human Factors) from Nanyang Technological University, Singapore in 2014.\\

\end{document}